\documentclass{iopjournal}
\usepackage{amsmath,amssymb,amsfonts}
\usepackage{dsfont}
\usepackage{graphicx}
\usepackage{tikz}
\newcommand{\abs}[1]{\left\vert#1\right\vert}
\newcommand{\mathbfh}[1]{\hat{\mathbf{#1}}}
\renewcommand{\imath}{\mathrm{i}}

\begin{document}

\articletype{Paper} %	 e.g. Paper, Letter, Topical Review...

\title{The power-spectrum tensor in steady-state systems
\\and its role in quantum friction}

\author{Francesco Intravaia$^{1,*}$\orcid{0000-0001-7993-4698}, and Kurt Busch$^{1,2}$\orcid{0000-0003-0076-8522}}

\affil{$^1$Institut f\"ur Physik, Humboldt-Universit\"at zu Berlin, Berlin, Germany}

\affil{$^2$Max-Born-Institut, Berlin, Germany}

\affil{$^*$Author to whom any correspondence should be addressed.}

\email{francesco.intravaia@physik.hu-berlin.de}


\begin{abstract}
We derive and classify properties of the power-spectrum tensor for systems in general steady-states, including stationary states not necessarily corresponding to equilibrium configurations. We establish a rigorous connection between the power-spectrum tensor and other quantities that characterize these systems, providing a systematic comparison with their equilibrium counterparts. As a physical application, we investigate the problem of quantum friction, describing the contactless quantum-electrodynamic drag acting on a particle moving in close proximity to material bodies at zero temperature. Specifically, we demonstrate how including additional information about the system's physical properties facilitates the derivation of more precise constraints on the power spectrum and its functional dependencies.
\end{abstract}

\section{Introduction} 
\label{Sec1}
Stochastic processes are omnipresent in engineering, physics and chemistry, so that
a detailed understanding of their properties lies at the heart of many applications
both, for fundamental and technological purposes.
Specifically, the power spectral density is one of the most relevant quantities for 
studying noisy time-stationary systems: It describes how the strength (commonly the 
power) of different components of stochastic processes is distributed across different frequencies~\cite{Van-Kampen92}. 
Within a noisy signal, this quantity allows for example to identify the dominant frequency-components, which, in turn, can be connected with the existence of  specific physical properties and the existence of underlying physical phenomena.
The concept of power spectral density can be generalized to multivariate or multidimensional stochastic processes where it becomes tensorial in nature and we speak of power-spectrum tensors.

Among the various fundamental relations involving the power spectrum, one of the most significant is the fluctuation-dissipation theorem (FDT)~\cite{Callen51,Kubo66}. For systems near equilibrium, this relation links the power-spectrum tensor to the susceptibility tensor. The latter characterizes the linear response to external perturbations, thereby capturing the dynamics of the system in the vicinity of its unperturbed state. Specifically, in the FDT both the power-spectrum and susceptibility tensors are defined as expectation values evaluated over the system's thermal equilibrium state. The theorem is related to the detailed balance between the different spontaneous energy-exchange processes that take place in the system~\cite{Talkner86,Speck06}. Probably 
one of the most remarkable feature of the FDT is its generality: As long as it is possible to define the system's power spectrum and susceptibility for the thermal equilibrium state, the details of the underlying microscopic dynamics do not matter. Whether the dynamics is linear or nonlinear, whether it is quantum or classical, whether it is Markovian or not~\cite{Talkner86,Van-Kampen92,Talkner09}, in the FDT the power spectrum is always proportional to the  susceptibility via a known factor that depends on temperature. Therefore,
it is not at all surprising that the theorem is applied in numerous and very 
diverse physical systems~\cite{Lifshitz56,Dzyaloshinskii61,Arcizet06,Intravaia11,Gieseler13,Gieseler14,Holubec22}, sometimes
even transcending its range of applicability.
In fact, despite the fact that the FDT is valid close to equilibrium, i.e. when the power spectrum and the susceptibility are evaluated at equilibrium, it has been considered also for systems with states being generically out of thermal equilibrium~\cite{Polder71,Dedkov02a,Antezza06,Manjavacas10a,Zhao12,Dadhichi23,Yu24} for which, technically speaking, it might become inaccurate. 
Arguments for the validity of this approach have typically been based on the 
so-called local thermal equilibrium (LTE) approximation. Within the LTE approximation, 
it is assumed that the correlation length and correlation time associated with the fluctuating dynamics 
are so short that the system's spatially separated components are (locally) in 
thermal equilibrium with their immediate surroundings~\cite{Intravaia16a,Intravaia19a,Reiche20c}. In a corse graining approach this usually allows for the definition of local intensive variables like the temperature which, contrary to a global equilibrium state, might now depend on the time and on the position within the system. 
The quality of this approximation depends strongly on the specific 
system, its physical composition and parameters~\cite{Samuelson70,Collins89} as well as on the involved length scales~\cite{Intravaia16a,Reiche20c}, thus rendering it at least questionable in a number
of cases. 

In this work, we examine the power-spectrum tensor and its fundamental properties, generalizing its connection to other physical quantities by evaluating expectation values over a nonequilibrium steady state (NESS) rather than an equilibrium state. In such stationary states, a system's stochastic properties are translation invariant with respect to time, despite the fact that the system can be far from equilibrium. In recent years, such stationary states have attracted particular attention~\cite{Oono98,Hatano01,Speck06,Sasa06,Chetrite08,Prost09,Gomez-Solano09,Gomez-Solano10,Seifert10} largely motivated by the research on fluctuation theorems more general than FDT and thermodynamic uncertainty relations~\cite{Talkner09,Esposito09,Seifert10,Campisi11,Fleming13,Rahav13,Barato15,Gingrich17,Wu20,Horowitz20,Dechant23}. These results establish connections between different physical quantities and lead to bounding inequalities with a broad range of validity, even beyond equilibrium.
To exist, the NESS must thus be sustained by one or several external agents such as externally maintained temperature gradients, chemical potential differences, or even mechanical drives that act on 
the system. In general, this can break the aforementioned detailed balance and, therefore, lead to violations of the FDT \cite{Speck06}. 

While we address nonequilibrium dynamics, our approach also includes the equilibrium case as 
a special limit. Specifically, we limit the assumptions about the microscopic dynamics of the system, thereby keeping the level 
of generality similar to that of the FDT. The stationarity of the NESS still allows us to draw a number of conclusions about the form and the mathematical properties of the power spectrum, allowing to obtain specific expressions that can be related to the physical characteristics of the underlying stochastic dynamics. These results can be useful to determine and analyze physical quantities relying on the power spectrum and to verify the consistency of expressions obtained using more detailed descriptions.

As an example for the validity of our results, we further discuss the properties of the power-spectrum tensor in the case of the nonconservative viscous 
motion of a particle in the quantum electromagnetic vacuum~\cite{Dedkov02a,Volokitin07,Milton16,Reiche22,Milton25}. Specifically, we address
quantum friction, i.e. the drag force acting on a neutral particle driven with constant velocity in the empty space
close to a generic arrangement of bodies.
Still keeping the microscopic dynamics unspecified, we show how additional information about the physics of the system 
strongly constrains the functional behavior and the structure of the power-spectrum tensor. 

The paper is organized as follows. In Sec.~\ref{genprop}, we consider the power-spectral density in the steady-state of a multidimensional stochastic 
process. We analyze its mathematical properties and how they relate to the physics of the system, establishing a connection to other quantities characterizing the system like its susceptibility.
In Sec.~\ref{qfriction}, we specify our analysis to the case of quantum friction and show how the thus increased physical information about 
the system allows us to derive additional relations. To confirm our results and illustrate 
the advantages as well as the simplifications introduced by our general approach, we consider here
an exactly solvable model for describing the internal dynamics of the particle.
Finally, in Sec.~\ref{conclusions} we summarize and discuss our findings.

\section{The power-spectrum tensor in steady-state systems}
\label{genprop}

\begin{figure}[h!]
    \vspace{-0.2cm}
  \centering
    \includegraphics[width=0.45\textwidth]{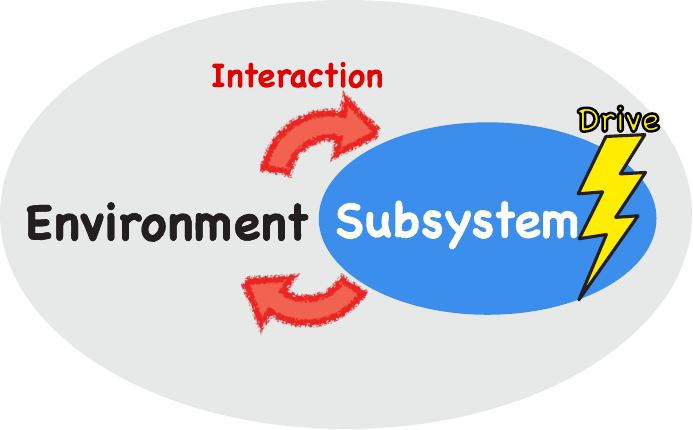}
    \hspace{0.03\textwidth}	
     \includegraphics[width=0.45\textwidth]{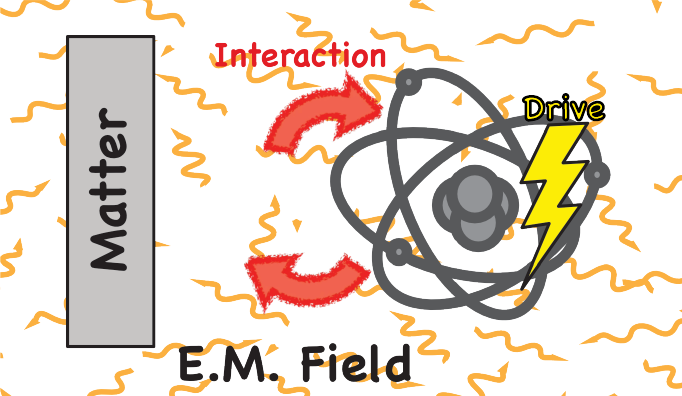}
     \begin{tikzpicture}[remember picture, overlay]
    \node at (-12.5,3.4) {{\bf a}};
     \node at (-6.1,3.4) {{\bf b}};
  \end{tikzpicture}
    \caption{{\bf a}: Sketch of the generic configuration analyzed in this work. A large, potentially driven quantum system composed of a subsystem interacting with an environment. The environment itself can be resulting from the interaction of different subsystems. {\bf b}: A possible physical implementation of this generic scheme: An externally driven atom coupled to a complex quantum environment, emerging from the self-consistent dynamics of the electromagnetic field in the presence of material bodies. The energy entering the system through the drive can be dissipated away in form of electromagnetic radiation and heat within the material, allowing the systems to reach a nonequilibrium steady state.
    	     \label{fig:scheme}
    	     }
		  \vspace{-0.2cm}
\end{figure}

We consider a generic, potentially driven quantum system. Its dynamics are generated by a Hamiltonian $\hat{H}_{t}$ that may explicitly depend on time. Our focus here is on a quantum system with a large number of degrees of freedom. It
is thought, in general, as being composed of a collection of interacting subsystems: On its own,      each of the quantum subsystems is \textit{open}, i.e. able to exchange energy and other physical quantities 
with the remaining parts of the system, representing its environment~\cite{Breuer02,Weiss08} (see Fig.~\ref{fig:scheme}{\bf a}). Therefore, the reduced dynamics of a subsystem is in general non-Hermitian~\cite{Breuer02,Weiss08,Bender07}. Contrary to typical approaches, however, we will refrain from deriving any master
equation, avoiding the Markov and other approximations which, depending on the specific circumstances, can be problematic for the description of quantum systems~\cite{Talkner86,Breuer16,Intravaia16}.
Our framework remains independent of microscopic details and assumes only that the system has reached a unique steady state $\hat{\rho}$, characterized by the condition $[\hat{\rho},\hat{H}_{t}]=0$.
Again, since energy can enter the system due to the drive, we would like to note explicitly that the time independent density operator $\hat{\rho}$ does not necessarily correspond to an equilibrium state and it can describe a system far from equilibrium.

Let $\hat{d}_{i}$ be the components of a Hermitian 
vector-operator $\mathbfh{d}$, describing physical quantities belonging to one of the interacting subsystems. The time-evolution in the Heisenberg picture of each component, $\hat{d}_{i}(t)$, is induced
according to $\hat{H}_{t}$ and we assume that all the components exhibit the same
time-reversal symmetry. 
Due to the interaction between the subsystems, the time-evolved operator  
$\hat{d}_{i}(t)$ encodes the global dynamics, reflecting the potentially high degree of entanglement between the system's various degrees of freedom.
For our purposes, we consider a three-dimensional vector but
generalizations to different dimensions are possible. The vector-operator $\mathbfh{d}$ either corresponds to a 
three-dimensional observable, or a quantity that characterizes a multivariate 
stochastic process as for example in three-particle systems. For simplicity, in the subsequent discussion, we will refer to
the former case but most of the results can be transferred to the latter case upon
adequate replacements of the relevant quantities. As a specific example, $\mathbfh{d}$ can describe a particle's electric dipole 
moment ($i=x,y,z$), interacting with a complex electromagnetic environment including objects comprised of dispersive and dissipative
materials (see Fig.~\ref{fig:scheme}{\bf b} and Sec.~\ref{qfriction}).

Since the system is stationary, the expectation value of $\mathbfh{d}(t)$ over 
$\hat{\rho}$ is constant. Without loss of generality, we assume it to be zero
%~\cite{Note1}.
\footnote{This condition can always be realized if we consider instead of 
	      $\hat{d}_{i}(t)$ the operator 
	      $\hat{\tilde d}_{i}(t)=\hat{d}_{i}(t)-\langle\hat{d}_{i}\rangle$.}, 
i.e. $\langle\hat{d}_{i}(t)\rangle\equiv\mathrm{tr}[\hat{d}_{i}(t)\hat{\rho}]=\mathrm{tr}[\hat{d}_{i}\hat{\rho}]=0$,
where `$\mathrm{tr}$' indicates the trace over a generic set of basis states
of the whole system's Hilbert space.
Given $\hat{d}_{i}(t)$ and $\hat{d}_{j}(t')$, 
one can define the corresponding correlation function 
\begin{equation}
\label{defcorr}
   C_{ij}(\tau) 
   \equiv 
   \langle\hat{d}_{i}(t)\hat{d}_{j}(t')\rangle
   =
   \mathrm{tr}\left[\hat{d}_{i}(t)\hat{d}_{j}(t')\hat{\rho}\right]~,
\end{equation}
which, for a stationary state, depends only on the time difference $\tau=t-t'$.
Owing to the fact that, in general, the involved operators do not commute, 
we may have complex-valued correlation functions even 
if the operators $\hat{d}_{i}$ are Hermitian~\footnote{In principle, we can avoid this by symmetrizing the product in the definition of $C_{ij}(\tau) $ (see for example \cite{Dalibard82,Landau80a}).}.
%~\cite{Note2} 
From Eq.~\eqref{defcorr} and the stationarity, we then further obtain that $C_{ij}^{*}(\tau)=C_{ji}(-\tau)$,
where the subscript `*' denotes complex conjugation. 
The set of correlation functions represents a correlation tensor $\underline{C}(\tau)$ so that we may
introduce a matrix notation for which we accordingly have that
\begin{equation}
\label{generalcrossig}
\underline{C}^{\dag}(\tau)=\underline{C}(-\tau).
\end{equation}
Here, the superscript `$\dagger$' denotes  the Hermitian conjugate. 

Relying on the Wiener-Khinchin theorem~\cite{Van-Kampen92}, the power-spectrum 
tensor $\underline{S}(\omega)$ is given by
\begin{equation}
\label{defPSD}
\underline{S}(\omega)=\int_{-\infty}^{\infty}\frac{\mathrm{d}\tau}{2\pi}\;\underline{C}(\tau) \; e^{\imath \omega \tau}.
\end{equation}
We start the analysis of this quantity by noticing that Eq.~\eqref{generalcrossig} implies $\underline{S}^{\dag}=\underline{S}$ so
that the power spectrum is necessarily Hermitian. As a consequence, we can 
always write
\begin{equation}
\label{decompositionS}
\underline{S}(\omega)=\underline{K}(\omega)+\boldsymbol{\Omega}(\omega)\cdot \mathbf{\underline{L}}~,
\end{equation}
where $\underline{K}(\omega)$ is a real symmetric matrix. The quantity 
$\boldsymbol{\Omega}(\omega)$ is a real three-dimensional vector so that 
$\boldsymbol{\Omega}\cdot \mathbf{\underline{L}}\equiv\sum_{k}\Omega_{k}\cdot \underline{L}_{k}$. 
Here $[\underline{L}_k]_{ij}=-\imath\epsilon_{kij}$ are the skew-symmetric 
(trace-less) matrix generators of the real group of three-dimensional rotations 
(SO(3) group; $\epsilon_{kij}$ indicates the Levi-Civita symbol). 

The first and second term on the r.h.s. of Eq.~\eqref{decompositionS} 
correspond, respectively, to the real and imaginary part of the power-spectrum 
tensor, thereby uniquely identifying their associated symmetry properties. 
The power-spectrum tensor $\underline{S}(\omega)$ is a 
positive semidefinite matrix (see App.~\ref{positivity}). 
Consequently $\underline{K}(\omega)=[\underline{S}(\omega)+\underline{S}^{\sf T}(\omega)]/2$
(the superscript `$\sf T$' denotes the transpose operation) 
is also positive semidefinite, since it is given by the sum of two positive semidefinite 
matrices.
Given that the orthogonal transformation that diagonalizes $\underline{K}(\omega)$ 
is an element of SO(3), we can find a basis where, in the decomposition of 
Eq.~\eqref{decompositionS}, $\underline{K}(\omega)$ is diagonal.
We can write the vector $\boldsymbol{\Omega}(\omega)$ as 
the Fourier transform of 
\begin{equation}
\boldsymbol{\Omega}(\tau)=\frac{\imath}{2}\langle  \mathbfh{d}(t)\times\mathbfh{d}(t')\rangle=\frac{\imath}{2}\langle  \mathbfh{d}(\tau)\times\mathbfh{d}(0)\rangle
\end{equation}
for which we obtain $\boldsymbol{\Omega}(-\tau)=\boldsymbol{\Omega}^{*}(\tau)$.
One can then interpret $\boldsymbol{\Omega}$ as a measure for the net rotation of $\mathbfh{d}(t)$ with 
respect to $\mathbfh{d}(t')$.
For classical systems, a similar quantity has been associated with 
the occurrence of nonequilibrium dynamics~\cite{Tomita74,Ohga23, Shiraishi23}, pointing to an analogous meaning for the part of $\underline{S}(\omega)$ that is proportional to
$\boldsymbol{\Omega}(\omega)$.

In connection with the power-spectrum tensor, it is also convenient to study the susceptibility tensor for the same stationary system. It is
defined as the Fourier transform of the corresponding linear response tensor $\underline{\alpha}(\tau)$. For our three-dimensional vectorial quantum-mechanical observable, the elements of the response tensor are given by (see App. \ref{positivity})
\begin{equation}
\label{polar}
 \alpha_{ij}(t-t')\equiv\theta(t-t')\langle \frac{\imath}{\hbar}[\hat{d}_{i}(t),\hat{d}_{j}(t')]\rangle~,
\end{equation}
where $\theta(t)$ denotes the Heaviside step function. 
Once again, due to stationarity, the linear response depends only on $\tau=t-t'$.
Given that $\underline{\alpha}(\tau)$ is real, its Fourier transform fulfills
what we call `crossing relation', i.e.  $\underline{\alpha}^{*}(\omega)=\underline{\alpha}(-\omega)$. 

As long as the Fourier transform exists, the Paley–Wiener theorem~\cite{Paley34} implies that the susceptibility $\underline{\alpha}(\omega)$ 
is analytical in the upper half of the complex frequency plane. As a consequence,
it fulfills the following properties:
Along the positive imaginary frequency axis, $\alpha_{ij}(\omega)$ is a real and 
monotonically decreasing function~\cite{Landau80a}
and its real and imaginary part are, respectively, even and odd functions of 
$\omega$. 
The exchange (absorption and emission) of energy connected with our vectorial quantum-mechanical observable due to a generic perturbation can be written in terms of 
\begin{equation}
\label{alphastrangei}
\underline{\alpha}_{\Im}(\omega)=\frac{\underline{\alpha}(\omega)-\underline{\alpha}^{\dag}(\omega)}{2\imath}~,
\end{equation}
which is proportional to the skew-Hermitian part of $\underline{\alpha}(\omega)$. 
Contrary to the linear susceptibility tensor, $\underline{\alpha}_{\Im}$ is a positive definite Hermitian matrix for $\omega>0$ (see App.~\ref{positivity})
and allows for a decomposition analogous to 
Eq.~\eqref{decompositionS}. 
Owing to the crossing relation, we have that $\underline{\alpha}_{\Im}(-\omega)=-\underline{\alpha}^{*}_{\Im}(\omega)=-\underline{\alpha}^{\sf T}_{\Im}(\omega)$, 
which implies that the real part of each matrix element is odd in $\omega$, while the corresponding imaginary part is even in the same variable. As a consequence, we also have that in general $\underline{\alpha}_{\Im}(0)\neq 0$ and is skew-symmetric. In connection with the previous properties, we can show that in general the symmetric part of despite $\underline{\alpha}(\imath\xi)$ must be positive definite. Indeed we have (see App. \ref{positivity})
\begin{equation}
\label{alphasymm}
\frac{\underline{\alpha}(\imath\xi)+\underline{\alpha}^{\sf T}(\imath\xi)}{2}=\frac{2}{\pi}\mathrm{Re}\int_{0}^{\infty}\mathrm{d}\omega \frac{\omega}{\omega^{2}+\xi^{2}}\underline{\alpha}_{\Im}(\omega)~.
\end{equation}
The elements of $\underline{\alpha}_{\Im}(\omega)$ are 
directly related to the commutator appearing in Eq.~\eqref{polar}. 
In particular, we obtain
\begin{equation}
\langle[\hat{d}_{i}(t),\hat{d}_{j}(t')]\rangle=\frac{\hbar}{\pi}\int_{-\infty}^{\infty}\mathrm{d}\omega\; [\underline{\alpha}_{\Im}(\omega)]_{ij}\; e^{-\imath \omega (t-t')}~,
\end{equation}
which can also be interpreted as a consequence of the analytical properties of the susceptibility tensor.

For the following considerations, it is convenient to introduce the matrix 
$\underline{\nu}(\omega)$
according to
\begin{equation}
\label{defnu}
\underline{\nu}(\omega)=\underline{S}(\omega)-\frac{\hbar}{2\pi}\underline{\alpha}_{\Im}(\omega)~.
\end{equation}
This matrix is Hermitian~\cite{Dalibard82,Dalibard84,Fleming13}, positive semidefinite (see App.~\ref{positivity}) and obeys the crossing relation $\underline{\nu}(\omega)=\underline{\nu}^{*}(-\omega)$. 
Its elements are the Fourier transforms of $\nu_{ij}(\tau)=\langle\{\hat{d}_{i}(t),\hat{d}_{j}(t')\}\rangle/2$, 
which are real-valued as they represent the expectation values of the anti-commutators 
of Hermitian operators. Consequently, the real and imaginary parts of $\nu_{ij}(\omega)$ are, respectively, even and odd functions of 
$\omega$.  In fact, the matrix $\underline{\nu}(\tau)$ is the 
symmetrized version~\cite{Dalibard82,Dalibard84} 
of the correlation tensor defined in Eq.~\eqref{defcorr}. From Eq.~\eqref{defnu}, taking the difference between $\underline{\nu}(\omega)$ and $\underline{\nu}^{*}(-\omega)$ and using the properties of the involved matrices, we find
\begin{equation}
\label{reverse}
\underline{S}^{*}(-\omega)=\underline{S}(\omega)-\frac{\hbar}{\pi}\underline{\alpha}_{\Im}(\omega)~.
\end{equation}
Since $\underline{S}$ and $\underline{\alpha}_{\Im}$ are, respectively, 
positive semidefinite and positive for $\omega>0$, we obtain that 
$\underline{S}(\omega)>\underline{S}^{*}(-\omega)=\underline{S}^{\sf T}(-\omega)$, 
where the inequality has to be interpreted in the sense of the Loewner order~\cite{Bhatia07}. 
The asymmetry of Eq.~\eqref{reverse} is shaped by
the quantum nature of the system. Indeed, classically one has that $\underline{C}(\tau) \in \mathds{R}$ implying 
that $\underline{S}^{*}(-\omega)=\underline{S}(\omega)$.

In the special case of the steady state being thermal equilibrium at the temperature
 $T$, i.e when the system is described by a time-independent Hamiltonian $\hat{H}_{0}$ and
\begin{equation}
\hat{\rho} = \frac{e^{-\beta\hat{H}_{0}}}{Z}, \quad Z=\mathrm{tr}[e^{-\beta\hat{H}_{0}}]
\end{equation}
where $\beta=1/[k_{B}T]$, the power spectrum and the susceptibility are 
connected by the expression 
\begin{equation}
\label{fdt}
\underline{S}(\omega)=\frac{\hbar}{\pi}\frac{\underline{\alpha}_{\Im}(\omega)}{1-e^{-\beta\hbar\omega}} 
\quad \text{(Equilibrium state)}.
\end{equation}
Indeed, this represents one of the possible forms of the FDT. 
Given that equilibrium is a special steady state, we can verify that Eq.~\eqref{fdt} must explicitly fulfill Eq.~\eqref{reverse}.
Specifically, here Eq.\eqref{reverse} can be seen as the manifestation of the
asymmetry introduced by quantum fluctuations in the absorption and emission processes characterizing the system dynamics. A concrete physical example of this phenomenon
is the spontaneous decay process characterizing an atom in the electromagnetic vacuum~\cite{Cohen-Tannoudji89}.

Whenever the Onsager reciprocity relations~\cite{Onsager31,Onsager31a,Casimir45,Landau80a} are valid, they
imply that, when averaged over the equilibrium state, the susceptibility $\underline{\alpha}$ must be symmetric so 
that $\underline{\alpha}_{\Im}\equiv\underline{\alpha}_{I}$ is real, where the subscript `$I$' 
indicates the element-wise imaginary party of the tensor. Considering Eq. \eqref{fdt},
one hat that the corresponding power-spectrum tensor is also real and, taking into account Eq. \eqref{decompositionS}, we can deduce that it is symmetric.
For reciprocal systems close to equilibrium, the Onsager reciprocity relations are thus equivalent to the condition 
$\boldsymbol{\Omega}=0$ which indicates that for an equilibrium state, despite possible (quantum) fluctuations, no net rotation 
of $\mathbfh{d}(\tau)$ around $\mathbfh{d}(0)$ can occur \cite{Tomita74,Ohga23, Shiraishi23}. 
Equivalently, from Eq.~\eqref{generalcrossig}, we have that the correlation tensor must 
fulfill the crossing relation (in time) $\underline{C}^{*}(\tau)=\underline{C}(-\tau)$ (the correlation tensor becomes symmetric).

For an arbitrary steady-state, we now write a relation which resembles 
Eq.~\eqref{fdt}:
\begin{equation}
\label{noneqDef}
\underline{S}(\omega)=\frac{\hbar}{\pi}\frac{ \underline{\alpha}_{\Im}(\omega)}{1-e^{-\beta\hbar\omega}}+\underline{J}(\omega)
\quad \text{(NESS)}~.
\end{equation}
Here, $\beta$ corresponds to the local equilibrium temperature 
of the system, often coinciding with that of its immediate surrounding. The matrix $\underline{J}(\omega)$ represents the correction to the 
FDT due to the system's possible far-from-equilibrium dynamics, going beyond the typical framework provided by linear-response theory. 
In fact, we can relate this correction to the net energy that flows between
the system and the environment in the NESS~\cite{Saito08,Harada05,Lippiello14,Intravaia14,Wu20,Reiche20c}.

The definition in Eq.~\eqref{noneqDef} allows us to render the question about the validity 
of the local thermal equilibrium approximation more precise, i.e. as a 
quantitative comparison of the impact of $\underline{J}(\omega)$ on the 
relevant physical quantities relative to that of the FDT-like term. As long 
as the contribution of $\underline{J}(\omega)$ can be neglected, the local 
thermal equilibrium approximation provides a viable approach. 
For a more detailed analysis, one often has to rely on the system's microscopic 
dynamics and on the specific physical quantity of interest (see also Sec.~\ref{3DOscillator}). 
Nonetheless, we can infer a number of general properties of  $\underline{J}(\omega)$. Indeed, 
since both $\underline{S}(\omega)$ and $\underline{\alpha}_{\Im}(\omega)$ 
are Hermitian, we must have that $\underline{J}(\omega)=\underline{J}^{\dag}(\omega)$. 
Upon employing the definition of $\underline{\nu}(\omega)$, we can write
\begin{equation}
\underline{J}(\omega)=\underline{\nu}(\omega)-\frac{\hbar}{2\pi}\coth\left(\frac{\beta\hbar \omega}{2}\right)\underline{\alpha}_{\Im}(\omega)~.
\end{equation}
Therefore, we can conclude that $\underline{J}(\omega)$ fulfills the crossing 
relation $\underline{J}(\omega)=\underline{J}^{*}(-\omega)=\underline{J}^{\sf T}(-\omega)$, 
highlighting that $\underline{J}(\tau)$ is real. 
The vanishing of $\underline{J}(\omega)$ can yet provide another version of the 
(equilibrium) FDT while $\underline{J}(\omega)\neq 0$ gives a measure of the 
violation of the detailed balance within NESS~\cite{Saito08,Harada05,Lippiello14,Intravaia14,Reiche20c}.
Specifically, for $T=0$ and $\omega>0$ we obtain
\begin{equation}
\underline{J}(\omega)=\underline{\nu}(\omega)-\frac{\hbar}{2\pi}\underline{\alpha}_{\Im}(\omega)=\underline{S}^{*}(-\omega)~,
\end{equation}
which allows us to state that in the limit of vanishing temperature $\underline{J}(\omega)$ 
must be positive semidefinite. In the same limit, we can write
\begin{align}
\label{FDTKK}
J_{ij}(\tau)=&\frac{1}{4\pi}\langle \{\hat{d}_{i}(\tau),\hat{d}_{j}(0)\}\rangle			
%\nonumber\\
%&
-\frac{1}{(2\pi)^{2}}\mathcal{P}\int_{-\infty}^{\infty}\mathrm{d}\tau'\;
\frac{\langle \imath [\hat{d}_{i}(\tau'),\hat{d}_{j}(0)]\rangle}{\tau'-\tau}~,
\end{align}
which specifically demonstrates how the conditions for the validity of the 
FDT imply the vanishing of $\underline{J}$. Indeed, in equilibrium, the correlations
$C_{ij}(\tau)=\langle\hat{d}_{i}(\tau)\hat{d}_{j}(0)\rangle=\langle \{\hat{d}_{i}(\tau),\hat{d}_{j}(0)\}\rangle/2-\imath \langle \imath [\hat{d}_{i}(\tau'),\hat{d}_{j}(0)]\rangle/2$ become analytical so that their 
real and imaginary part must fulfill the Kramer-Kronig relations~\cite{Pottier01}, 
for which the terms on the r.h.s. of Eq.~\eqref{FDTKK} cancel to zero \footnote{
More precisely, we would like to recall that the correlation tensor
$\underline{C}(\tau)$, in the equilibrium state at a generic temperature $T$, satisfies the following two additional properties: 
The elements of the matrix $\underline{C}^{*}(\tau)$ are analytical 
functions within the stripe $0< \tau_{I} <\hbar\beta$ in the upper
half  part of the complex $\tau$-plane~\cite{Martin59}; 
In addition, in an equilibrium state $\underline{C}(\tau)$ fulfills the Kubo-Martin-Schwinger condition, i.e. 
$\underline{C}^{*}(\tau)=\underline{C}(\tau-\imath\hbar\beta)$
~\cite{Kubo57,Kubo66,Pottier01,Martin59,Breuer02}.}.

The considerations above illustrate that, in addition to possibly having $\boldsymbol{\Omega}\neq 0$, the system's quantum 
nonequilibrium steady state is in general characterized by $\underline{J}\neq 0$. Although both 
quantities are connected to the breaking of the detailed balance occurring at equilibrium, the underlying concepts
they are addressing are, in general, not identical. As we will see below, there are indeed situations for which $\boldsymbol{\Omega}= 0$
and $\underline{J}\neq 0$. 

\section{Viscous motion through the electromagnetic vacuum}
\label{qfriction}

\begin{figure}[h!]
    \vspace{-0.2cm}
  \centering
    \includegraphics[width=0.45\textwidth]{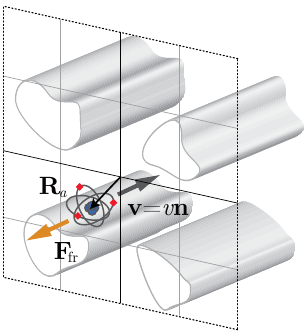}	
    \caption{Schematic representation of a particle moving in vacuum close to an arrangement of different, macroscopic objects. The particle moves on a straight line with direction $\mathbf{n}$, which coincides with the direction of translational 
    	     invariance of the arrangement of macroscopic bodies. In this 
    	     setup, the quantum fluctuations of the electromagnetic field behave as 
    	     a viscous medium and the particle's motion is affected by a corresponding
    	     drag force. An external force acting on the particle's center of mass 
    	     (not shown) keeps the motion at constant velocity $v$. The power generated by the external drive is dissipated in the electromagnetic plus dissipative-matter environment~\cite{Intravaia15,Reiche20c}.
    	     \label{fig:generic-setup}
    	     }
		  \vspace{-0.2cm}
\end{figure}

As soon as the system's physical realization becomes more specific, its properties and symmetries 
allow us to draw further conclusions on its stochastic dynamics. In this section we illustrate this aspect within the framework
of quantum electrodynamics, where the quantum fluctuations pervading the system are those 
connected with the absorptive, dissipative and dispersive processes characterizing light-matter interaction. 
As a concrete application, we consider an electrically neutral particle (e.g. an atom) which can move in vacuum
with nonrelativistic velocity along a prescribed trajectory. We further assume
that the trajectory of its center of mass is classical (usually a good approximation for particles with relatively large masses or velocities) and parallel to the translationally invariant direction of 
an arbitrary arrangement of electrically neutral material bodies, see Fig.~\ref{fig:generic-setup}. 
We denote the direction of motion with the unit vector $\mathbf{n}$ and we
use the vector $\mathbf{R}_{a}$ to identify the particle's lateral position. 
The material bodies are comprised of linear, reciprocal and passive materials 
which we further assume to be nonmagnetic, dissipative, possibly spatially 
dispersive and, in absence of the particle, in thermal equilibrium with the surrounding radiation.
%%%%%%%%%%%%%%
%%%%%%%%%%%%%%
Despite the system being electrically neutral, when the particle moves along its trajectory, it interacts with the fluctuations of the electromagnetic field which, by the presence of the material bodies, are modified 
relative to vacuum. As a result of this interaction, a frictional force acts on the particle~\cite{Dedkov02a,Volokitin07,Reiche22} (see also below)
that tends to bring the particle to rest with respect to the privileged reference frame defined by the arrangement of the material bodies. 
Consequently, an external force acting on center of mass of the subsystem particle is required to maintain its motion within the quantum environment resulting from the interaction of dissipative matter and the electromagnetic radiation.
When an external drive and the frictional force balance, the system reaches 
a nonequilibrium steady-state characterized by the particle moving at constant speed $v$. In this state the input power, due to the drive acting on the atom, is exactly dissipated in the environment~\cite{Intravaia15,Reiche20c}.

The electromagnetic viscosity experienced by the particle~\cite{Oelschlager22} is a manifestation of the fluctuating and dissipative nature of its quantum environment. Due to the existence of quantum fluctuations, the drag has the peculiarity to exist even at zero
temperature, where it is usually referred to as quantum friction.
Within the standard framework, the stationary fluctuating dynamics of the particle is usually described by its three-dimensional 
electric dipole (vector-)operator $\mathbfh{d}(t)$. In the above-discussed nonequilibrium 
steady-state, the corresponding correlation tensor, $ \underline{C}(\tau,v)$, and the 
power-spectrum tensor, $\underline{S}(\omega,v)$, explicitly depend on the particle velocity
$v$ which thus plays the role of a control variable
of our nonequilibrium configuration (naturally, $v=0$ indicates in this case the equilibrium state)~\cite{Rahav13}. 
In general, these tensors also depend on the lateral position $\mathbf{R}_{a}$ but in order to
avoid notational clutter, in the subsequent discussions, we suppress the dependence on $\mathbf{R}_{a}$ unless explicitly required.
Clearly, for any value of $v$ and $\mathbf{R}_{a}$ the power-spectrum tensor and the related quantities 
must fulfill the relations discussed in Sec.~\ref{genprop}.

In the NESS, the particle's dipole does not only viscously move with velocity $v$ but, due to the interaction with its quantum electromagnetic environment, it
can also gain and loose angular momentum~\cite{Intravaia19a,Reiche20c,Reiche20a}.
However, owing to the entire system's symmetry and its intrinsic dissipative behavior, in the stationary state the particle cannot have 
a net angular momentum component with the direction parallel to $\mathbf{n}$.
As a physical consequence, the dipole cannot feature a net averaged rotation around the direction of motion and
Eq.~\eqref{decompositionS} takes the form
\begin{equation}
\label{decompositionS-QF}
\underline{S}(\omega,v)=\underline{K}(\omega,v)+\boldsymbol{\Omega}_{\perp}(\omega,v)\cdot \mathbf{\underline{L}}_{\perp}~,
\end{equation}
where $\perp$ indicates a direction orthogonal to $\mathbf{n}$. 

Equation \eqref{decompositionS-QF} can be used as starting point for discussing some general aspects about
the tensorial structure of $\underline{S}(\omega,v)$ and its functional dependence on $v$.
Indeed, given that for $v=0$ our system is in equilibrium, reciprocal and, therefore, the power spectrum is 
real and symmetric, we can already conclude that $\boldsymbol{\Omega}_{\perp}(\omega,0)=0$.
Since the consituent materials are reciprocal, we have 
\begin{equation}
\label{symmetryTransf}
\underline{S}(\omega,v)=\underline{M}_{n}\underline{S}(\omega,-v)\underline{M}^{-1}_{n}~,
\end{equation}
which is a consequence of the system's mirror-symmetry.
Here, we denote with $\underline{M}_{n}=\underline{\mathds{1}}-2\mathbf{n}\mathbf{n}^{\sf T}$ 
the parity transformation that describes the reflection with respect to the plane orthogonal 
to $\mathbf{n}$ ($\underline{\mathds{1}}$ is the identity matrix). The change in sign of the 
velocity results from applying $\underline{M}_{n}$ to the velocity vector $\mathbf{v}=v\mathbf{n}$, 
i.e. $\underline{M}_{n}\mathbf{v}=-v\mathbf{n}$. However, the same transformation leaves
$\mathbf{R}_{a}$ and all (proper) vectors parallel to the plane of symmetry unchanged.
Further, we find it convenient to decompose the real matrix $\underline{K}$ as 
\begin{equation}
\label{decompositionK}
\underline{K}(\omega,v)=\underline{\Delta}(\omega,v)+\boldsymbol{\chi}(\omega,v)\cdot\mathbf{\underline{\Lambda}}~,
\end{equation}
where $\underline{\Delta}$ is diagonal and $\boldsymbol{\chi}$ is a three-dimensional real-valued 
vector. The matrices $[\underline{\Lambda}_k]_{ij}=|\epsilon_{kij}|$ are symmetric, traceless 
and trace-orthogonal to $\underline{L}_{i}$ and $\underline{\Delta}$. In particular, due to 
their connection with the Gell-Mann matrices~\cite{Fuchs03}, the decomposition over all the previous matrices 
is unique. 

If we denote with $\underline{\Lambda}_n=\mathbf{n}\cdot\mathbf{\underline{\Lambda}}$ the matrix 
that corresponds to the direction $\mathbf{n}$, we obtain that $\underline{\Lambda}_n$ and 
$\underline{\Delta}$ are invariant with respect to $\underline{M}_{n}$. For the remaining 
matrices, we find the transformations  $\underline{M}_{n}\boldsymbol{\underline{\Lambda}}_{\perp}\underline{M}^{-1}_{n}=-\boldsymbol{\underline{\Lambda}}_{\perp}$ and $\underline{M}_{n}\mathbf{\underline{L}}_{\perp}\underline{M}^{-1}_{n}=-\mathbf{\underline{L}}_{\perp}$.
Due to the uniqueness of the decomposition, we can thus conclude from Eq.~\eqref{symmetryTransf} 
that $\underline{\Delta}(\omega,v)$ and $\chi_{n}(\omega,v)$ are even in the velocity $v$, while 
$\boldsymbol{\chi}_{\perp}(\omega,v)$ and $\boldsymbol{\Omega}_{\perp}(\omega,v)$ are odd
functions of $v$.  
Taking into account that $\underline{L}^{\sf T}_{i}=-\underline{L}_{i}$ we can further write 
(see App.~\ref{appSymm})
\begin{align}
\label{transposeIdentity}
\underline{S}(\omega,-v)=\underline{S}^{\sf T}(\omega,v)-2\boldsymbol{\chi}_{\perp}(\omega,v)\cdot\mathbf{\underline{\Lambda}}_{\perp}~,
\end{align}  
which represents the formal counterpart of Eq.~\eqref{reverse} for the variable $v$.
By proceeding in an analogous way or by exploiting their connection to the power-spectrum tensor, we can derive similar relations for the matrices
discussed in the previous section. 
Specifically, from Eq.~\eqref{reverse} we obtain
\begin{equation}
\label{alphstsp}
\underline{\alpha}_{\Im}(\omega,-v)=\underline{\alpha}_{\Im}^{\sf T}(\omega,v)-2\pi\frac{\boldsymbol{\chi}_{\perp}(\omega,v)-\boldsymbol{\chi}_{\perp}(-\omega,v)}{\hbar}\cdot\mathbf{\underline{\Lambda}}_{\perp}~.
\end{equation}
An interesting special case for this relation occurs if the expression for the susceptibility of the quantum dipole operator is identical to that of its classical counterpart or does not depend on temperature. Under these circumstances, the last term on the r.h.s. of Eq.~\eqref{alphstsp} must always coincide with its classical limit. Since we already know that, classically, $\underline{S}^{*}(-\omega)=\underline{S}(\omega)$, we directly infer that $\boldsymbol{\chi}_{\perp}(-\omega,v)=\boldsymbol{\chi}_{\perp}(\omega,v)$. Therefore, the last term of Eq.~\eqref{alphstsp} vanishes, leading to the identity $\underline{\alpha}_{\Im}(\omega,-v)=\underline{\alpha}_{\Im}^{\sf T}(\omega,v)$. Due to its analyticity and the Kramers-Kronig relations, this means that when the susceptibility has the same form within the quantum and classical description, it must fulfill $\underline{\alpha}(\omega,-v)=\underline{\alpha}^{\sf T}(\omega,v)$.
This is tantamount to stating that, in this case, the system fulfills Onsager–Casimir-like 
relations where the atom's speed $v$ takes the role of the external 
bias parameter~\cite{Asadchy20}.

If the system exhibits an additional mirror-symmetry with respect to the reflection $\underline{M}_{t}$, where $\mathbf{t}$ 
is a unit vector orthogonal to $\mathbf{n}$, then we can show along the lines discussed above 
that $\boldsymbol{\chi}=\chi_{t}\mathbf{t}$ and $\boldsymbol{\Omega}_{\perp}=\Omega_{t}\mathbf{t}$  
(see App.~\ref{appSymm}). 
However, while the symmetry described by $\underline{M}_{n}$ is independent of the transverse
location $\mathbf{R}_{a}$, the symmetry associated with $\underline{M}_{t}$ can, in general, 
only exist when $\mathbf{R}_{a}$ is located within (or parallel to) the symmetry plane. In 
other words, conclusions about $\underline{M}_{t}$ do depend on the transverse position of 
the particle.
Similarly, if an additional mirror-symmetry along the direction $\mathbf{n}\times \mathbf{t}$ 
exists, this necessarily implies that, when the particle moves along the associated symmetry 
axis (located at the intersection of the two corresponding symmetry planes), the power-spectral 
tensor is diagonal, suppressing any possible rotation of the dipole operator
~\cite{Intravaia19a,Reiche20a,Oelschlager22}. As mentioned above, although in this case 
the particle is characterized by a specific position $\mathbf{R}_{a}$, this is an example of nonequilibrium
dynamics for which  $\boldsymbol{\Omega}=0$.
In addition, we would like to point out that the above reasoning essentially identifies a specific physical reference frame with one 
of the axes coinciding with $\mathbf{n}$, within which the power-spectrum tensor has a 
diagonal form. Indeed, while a unitary transformation that diagonalizes the power-spectrum 
tensor always exists, this transformation does, in general, not correspond to a physical rotation.

So far, we have assumed that the three components of the dipole operator can be varying 
in time independent of each other. Some of the above conclusions are modified for a particle's 
model in which the dynamics along the three axes is correlated by construction. This occurs, for instance, 
if we consider a common model in quantum optics where the particle's dipole operator is 
modeled as $\mathbfh{d}=\mathbf{d}\hat{\sigma}$, where $\mathbf{d}$ is a constant 
three-dimensional real-valued vector and $\hat{\sigma}$ is a scalar operator -- usually 
a Pauli operator for a simple two-state system~\cite{Scheel09,Intravaia16a,Oelschlager18,Farias20,Viotti21} 
or a 1D-harmonic oscillator~\cite{Intravaia16,Farias19}. 
In this case, the correlation tensor is always proportional to the real and symmetric 
dyadic $\mathbf{d}\mathbf{d}^{\sf T}$, i.e.
\begin{equation}
\underline{C}(\tau,v)
=
\mathbf{d}\mathbf{d}^{\sf T}\langle \hat{\sigma}(t+\tau)\hat{\sigma}(t+\tau)\rangle
=
\mathbf{d}\mathbf{d}^{\sf T}C_{\sigma}(\tau,v)~.
\end{equation}
As a consequence, we find $\underline{C}(\tau,v)=\underline{C}^{\sf T}(\tau,v)$ and 
from the general properties of the correlation function we obtain $C_{\sigma}(\tau,v)=C^{*}_{\sigma}(-\tau,v)$. 
The corresponding power-spectrum tensor $\underline{S}(\omega,v)=\mathbf{d}\mathbf{d}^{\sf T}\Sigma(\omega,v)$  
is real and symmetric, implying that within this model $\boldsymbol{\Omega}$ vanishes
exactly, irrespective of the particle's position.
From the physical point of view, this is equivalent to the inability of the dipole vector 
to rotate around its initial value, which is consistent with having the real vector $\mathbf{d}$ being time-independent. 
This means that, within this model, the particle cannot absorb or emit angular momentum
~\cite{Intravaia19a}.  
Using this model, we obtain from symmetry considerations that $\Sigma(\omega,v)$ must 
be an even function of $v$. We can see this most easily by considering the case where 
the dipole operator is oriented orthogonal to the direction of motion, i.e. 
$\mathbf{d}\equiv \mathbf{d}_{\perp}$. In this case, Eq.~\eqref{symmetryTransf} in
conjunction with fact that the matrix $\mathbf{d}_{\perp}\mathbf{d}_{\perp}^{\sf T}$ 
is invariant with respect to $\underline{M}_{n}$, we find that $\Sigma(\omega,v)=\Sigma(\omega,-v)$.

\subsection{The power spectrum in quantum friction}

In order to see how the information above is relevant for understanding the physics of the system, 
let us consider the expression for the quantum frictional force. 
If the system is at zero temperature, earlier work
~\cite{Intravaia14,Intravaia16,Intravaia19a,Reiche20a} 
has shown that in the NESS the frictional force 
is oriented along the direction of translational invariance, i.e. 
$\mathbf{F}_{\rm fr}=F_{\rm fr}\mathbf{n}$ (see Fig.~\ref{fig:generic-setup}), and 
is of the form
\begin{equation}
   F_{\rm fr}= - 2\int_{0}^{\infty}\hspace{-2mm} \mathrm{d}\omega
	         \int \frac{\mathrm{d} q}{2\pi} \, q \; \mathrm{Tr}\left[\underline{S}^{\sf T}(-\omega^{-}_{q},v)
			                                      \underline{G}_{\Im}(q,\mathbf{R}_{a}, \omega)\right].
\label{totalForce}
\end{equation} 
where we indicate with `$\mathrm{Tr}$' the trace over the tensors' indices, while $q$ is the component of the electromagnetic field's wave vector along the direction of motion and 
$\omega^{\pm}_{q} = \omega\pm q v$ is the Doppler-shifted frequency. 
In addition to the dipole's power-spectrum tensor $\underline{S}$, the 
interaction is described in terms of the electromagnetic Green tensor 
$\underline{G}$: Physically, it represents the linear 
electromagnetic susceptibility tensor of the system without the particle, i.e. it exclusively characterizes the (electromagnetic) environment with which the particle interacts. For the aforementioned arrangement of the material bodies that comprise the system, $\underline{G}$ can be
in principle obtained by solving the corresponding Maxwell equations with a source term \cite{Intravaia16,Kristensen23}.

The physical interpretation of Eq.~\eqref{totalForce} identifies the l.h.s. as the net total momentum per unit of time that is transferred to or from 
the particle during the stochastic dynamics connected with the underlying 
absorption and emission processes characterizing the interaction~\cite{Maghrebi13,Intravaia15,Intravaia19a,Oelschlager22}. 
Specifically, $\underline{G}_{\Im}=[\underline{G}-\underline{G}^{\dag}]/(2\imath)$ in Eq.~\eqref{totalForce} is connected with the electromagnetic density of states~\cite{Joulain05}. Moreover, in complete analogy to $\alpha_{\Im}$, we may relate the matrix $\underline{G}_{\Im}$
to the exchange of energy (dissipation) with the electromagnetic 
environment that surrounds the particle and is consequently positive 
semidefinite for $\omega>0$ (see also the discussion around 
Eq.~\eqref{eq:flucdisE0}).
Technically, the Green tensor represents a propagator of the electromagnetic
field and as such usually depends on two spatial arguments~\cite{Jackson75,Morse53},
a source position and a detection position. However, since in all our 
expressions we require the back-action of the electromagnetic field 
emitted by the particle onto itself (via scattering by the material bodies), source and detection lateral positions coincide and
we retain in the Green tensor only one spatial argument. In addition, given
the system's translational invariance along the particle trajectory, we carry out 
a Fourier transform, so that the Green tensor acquires a mixed 
Fourier- and real-space dependence with regards to the positional arguments 
and, of course, features a dependence on frequency, i.e. 
$\underline{G}(q, \mathbf{R}_{a}, \omega)$.
This expression fulfills the identities
\begin{subequations}
\label{propgi}
\begin{align}
\underline{G}_{\Im}(-q,\mathbf{R}_{a}, \omega)&= \underline{G}^{\sf T}_{\Im}(q,\mathbf{R}_{a}, \omega)\label{trps}~,
\\
\underline{G}_{\Im}(q,\mathbf{R}_{a}, -\omega)&= -\underline{G}_{\Im}(q,\mathbf{R}_{a}, \omega)~,
\end{align}
\end{subequations}
which we may, respecitively, view as consequences of the system's 
reciprocity in the real space and stationarity in time.

Furthermore, since $\underline{G}_{\Im}$ is Hermitian, we can decompose it
into the sum of a diagonal matrix and a linear combination with real-valued
coefficients of the matrices $\underline{L}_{i}$ and $\underline{\Lambda}_{i}$. 
From the reciprocity of the materials, given the mirror symmetry of the 
macroscopic bodies, we then have that
\begin{equation}
\label{symmG}
\underline{G}_{\Im}(q,\mathbf{R}_{a}, \omega)=\underline{M}_{n}\underline{G}_{\Im}(-q,\mathbf{R}_{a}, \omega)\underline{M}^{-1}_{n}~.
\end{equation}
This would lead to an expression of the same form as provided in Eq.~\eqref{transposeIdentity}. 
However, due to the (system-specific) condition in Eq.~\eqref{trps}, 
we can go one step further and find that the coefficients associated with 
$\boldsymbol{\underline{\Lambda}}_{\perp}$ and $\underline{L}_{n}$ actually 
vanish (see App.~\ref{appSymm}) so that we can write 
\begin{equation}
\label{split}
\underline{G}_{\Im}=\underline{D}+c_{n}\underline{\Lambda}_{n}+\mathbf{s}_{\perp}\cdot\mathbf{\underline{L}}_{\perp}~.
\end{equation}
Here, the matrix $\underline{D}(q,\mathbf{R}_{a}, \omega)$ is real-valued and 
diagonal. Both, the matrix $\underline{D}(q,\mathbf{R}_{a}, \omega)$ and the function $c_{n}(q,\mathbf{R}_{a}, \omega)$ are even in $q$.
The real vector $\mathbf{s}_{\perp}(q,\mathbf{R}_{a}, \omega)$ is odd in $q$ 
and can be related to a spin-dependent part of the electromagnetic density 
of states
~\cite{Intravaia19a,Mandel95}, 
thereby describing the so-called spin-momentum locking of light
~\cite{Bliokh15,Lodahl17,OShea13,Sayrin15,Gong18}. 
Analogous to the power-spectrum tensor, if the arrangement of material bodies 
is also mirror-symmetric with respect to $\underline{M}_{t}$ and $\mathbf{R}_{a}$ 
lies in (or is parallel to) the plane of symmetry, we obtain $c_{n}=0$ and 
$\mathbf{s}_{\perp}=s_{t}\mathbf{t}$. By the same token, if a symmetry axis 
exist and $\mathbf{R}_{a}$ lies on this axis, $\underline{G}_{\Im}$ is diagonal.

Combining the results on $\underline{G}_{\Im}$ with those on the power-spectrum 
tensor, we can infer a number of properties of the frictional force. When the 
three components of the dipole operator can have independent dynamics and upon 
considering Eqs.~\eqref{decompositionS-QF}, \eqref{decompositionK} and \eqref{split}, 
we can write the trace in the expression for the frictional force as
\begin{subequations}
\label{traces}
\begin{align}
\label{tr1}
\mathrm{Tr}\left[\underline{S}^{\sf T}  \underline{G}_{\Im}\right]= 
\mathrm{Tr}\left[\underline{\Delta}\underline{D}\right]+2\chi_{n}c_{n}
-2\boldsymbol{\Omega}_{\perp}\cdot\mathbf{s}_{\perp}.
\end{align}
Here, we have utilized that $\mathrm{Tr}[\underline{L}_{i}\underline{L}_{j}]=\mathrm{Tr}[\underline{\Lambda}_{i}\underline{\Lambda}_{j}]=2\delta_{ij}$ 
and 
$\mathrm{Tr}[\underline{L}_{i}\underline{\Lambda}_{j}]=0$. 
For the case where $\mathbfh{d}(t)=\mathbf{d}\sigma(t)$ we find instead 
\begin{align}
\label{tr2}
\mathrm{Tr}\left[\underline{S}^{\sf T}  \underline{G}_{\Im}\right]
&= \mathbf{d}\cdot\underline{G}_{\Im}\cdot \mathbf{d}  \;\Sigma
=\mathbf{d}\cdot\left(\underline{D}+c_{n}\Lambda_{n}\right)\cdot \mathbf{d}\;\Sigma
\nonumber\\
&\stackrel{\text{average}}{=} \frac{d^{2}}{3}\mathrm{Tr}\left[ \underline{G}_{\Im}\right]\;\Sigma = \frac{d^{2}}{3}\mathrm{Tr}\left[\underline{D}\right]\;\Sigma~.
\end{align}
\end{subequations}
The last line in the above equations corresponds to the result of an average 
over all possible directions of the vector $\mathbf{d}$ for which the power-spectrum tensor 
becomes proportional to $\underline{\mathds{1}}d^{2}/3$ ($d^{2}=\mathbf{d}\cdot\mathbf{d}$)
~\cite{Intravaia16a,Reiche17,Oelschlager18}.
When comparing the two results, we notice, as expected, that the main difference lies in the contribution
arising from $\boldsymbol{\Omega}_{\perp}$, which is absent in the second model. Moreover, when
$\boldsymbol{\Omega}_{\perp}$ and $\mathbf{s}_{\perp}$ have the same sign, the corresponding
contribution tends to reduce the frictional force. This confirms and generalizes earlier results in Refs.~\cite{Intravaia19a,Reiche20a}, showing that they are not
tied to the specific model for the dipole's dynamics used there (see Sec.~\ref{3DOscillator}).
The remaining contributions in Eqs.~\eqref{traces} are similar in both models although not identical. Specifically, the terms proportional to 
$c_{n}$ directly show how the system's symmetry (or better the lack thereof) affects the frictional force. More precisely, as
soon as the system is mirror-symmetric with respect to a plane containing the particle's trajectory -- for example when
the particle moves above a planar structure -- the terms proportional to
$c_{n}$ in Eqs.~\eqref{traces} vanish. Also, averaging over the dipole direction in the second model, washes out this contribution which does not appear in the last line of Eq.~\eqref{tr2}.

Another interesting consequence is that the results in Eqs.~\eqref{traces} are 
both equivalent to the identity
\begin{equation}
\label{oddinvidentity}
\mathrm{Tr}\left[\underline{S}^{\sf T}(-\omega^{+}_{q},-v)\underline{G}_{\Im}(q,\mathbf{R}_{a}, \omega)\right]
=\mathrm{Tr}\left[\underline{S}(-\omega^{+}_{q},v)\underline{G}_{\Im}(q,\mathbf{R}_{a}, \omega)\right]~.
\end{equation}
When used in Eq.~\eqref{totalForce} in conjunction with Eq.~\eqref{trps} 
and after changing $q \to -q$, the previous expression allows us to show that, in general, the frictional 
force must always reverse its sign if $v\to -v$, i.e. $F_{\rm fr}(-v)=-F_{\rm fr}(v)$. In other words, independently of the model for the dipole's dynamics, $F_{\rm fr}(v)$ is an odd function of the velocity. 
This is what is physically expected for a frictional interaction and has been confirmed by calculations with specific models \cite{Kyasov02,Pieplow13,Volokitin14,Intravaia14,Intravaia15,Oelschlager22}. We want to underline, however, that the mathematical properties justifying this behavior are less trivial than one could intuitively presume. Indeed, reversing the logic and imposing that $F_{\rm fr}(v)$ is odd with respect to the velocity one would obtain Eq.~\eqref{oddinvidentity}, which can erroneously suggest that $\underline{S}^{\sf T}(\omega,-v)=\underline{S}(\omega,v)$, in contradiction with Eq.~\eqref{transposeIdentity}.

\subsection{An illustrative example}
\label{3DOscillator}

Thus far, we have refrained from assuming particular microscopic dynamics for the electric dipole operator. To illustrate the validity and physical implications of our results, we now introduce a tractable model that admits an exact solution for the equations of motion.
Specifically, we assume that the particle moving in vacuum parallel to a translational invariant arrangement of bodies is an atom.
Given that the interaction between the atomic internal degrees of freedom and 
the electromagnetic fluctuations is typically weak, it is reasonable to 
describe the dipole operator in term of linear system associated with a 
single frequency $\omega_a$. Physically, $\omega_a$ would correspond to 
the transition frequency between the highest occupied orbital to the lowest 
unoccupied one.
Within this linear approximation, we are going to represent $\mathbfh{d}$ 
in terms of a 3D isotropic quantum-mechanical oscillator
~\cite{Hu92,Intravaia14} 
which fulfills the following equation of motion
\begin{equation}
  \label{eq:deom}
  \left(\partial_t^2+\omega_a^2\right)
  \hat{\mathbf{d}}(t)
  =
  \alpha_0 \omega_a^2
  \hat{\mathbf{E}}(\mathbf{r}_a(t),t)~ .
\end{equation}
In this expression, $\alpha_0$ is a constant which essentially assesses 
the strength of the coupling between the atom and the electromagnetic 
radiation. Specifically, it can be identified with the static polarizability 
associated with the aforementioned atomic transition 
~\cite{Intravaia16,Intravaia19a}. 

The r.h.s. of Eq.~\eqref{eq:deom} denotes the total fluctuating electric 
field that drives the dipole and it is evaluated at the position 
$\mathbf{r}_a(t)=x_{a}(t)\mathbf{n}+\mathbf{R}_{a}$ which represents the 
trajectory of the atom's center of mass.
We find it convenient to decompose the electric field operator into two 
contributions, i.e. $\mathbfh{E}(\mathbf{r},t)=\mathbfh{E}_{0}(\mathbf{r},t)+\mathbfh{E}_{\rm s}(\mathbf{r},t)$,
where the first terms, $\mathbfh{E}_0(\mathbf{r},t)$, represents the 
electric field which would exist without the atom and we assume it
to be thermalized. Consequently, this field obeys the fluctuation-dissipation 
theorem~\cite{Kubo66,Intravaia16,Reiche20a}, 
and we can write
\begin{equation}
  \label{eq:flucdisE0}
  \langle
  \hat{\mathbf{E}}_0(q;\mathbf{R}_a ;\omega)
  \hat{\mathbf{E}}^{\mathsf{T}}_0(\tilde{q}, \mathbf{R}_a ;\tilde{\omega})
  \rangle
  = 
  \frac{8\pi^2\hbar}{1-e^{-\beta\hbar\omega}}
  \underline{G}_{\Im}(q,\mathbf{R}_a,\omega)
  \delta(\omega+\tilde{\omega})\delta(q+\tilde{q})
  \,,
\end{equation}
where $\delta(\cdot)$ is the Dirac function. 
The second contribution,  $\mathbfh{E}_{\rm s}(\mathbf{r},t)$, describes the 
radiation induced by the atom. It corresponds to the special solution of 
Maxwell equations having the dipole as source and, as described above, 
it can be written as
\begin{align}
  \label{eq:efield}
  \mathbfh{E}_{\rm s}(\mathbf{r},t)
   & =
  \int\limits \mathrm{d}t'\,
  \underline{G}(\mathbf{r},\mathbf{r}_a(t'),t-t')\,
  \hat{\mathbf{d}}(t')~,
\end{align}
where $ \underline{G}$ is the Green tensor, represented in this case in time and space.

When the NESS is achieved, we have that $x_{a}(t)\sim v t$~\cite{Intravaia16} and combining 
the above expression with Eq.~\eqref{eq:deom}, we Fourier transform into 
the frequency domain and can solve the system self-consistently up to all 
orders in the coupling $\alpha_0$
~\cite{Reiche20a}. 
The stationary solution of the dipole's dynamics takes the form
\begin{equation}
\label{dss}
\mathbfh{d}(\omega)=\int \frac{\mathrm{d} q}{2\pi}\underline{\alpha}(\omega,v)\cdot\mathbfh{E}_{0}(q,\mathbf{R}_{a}; \omega_{q}^{+})~,
\end{equation}
where $\underline{\alpha}(\omega, v)$ is the steady-state dressed atomic 
polarizability defined as
  \begin{align}
     \label{Eq:PolDressed}
    \underline{\alpha}(\omega, v)
     & =
    \left[
      \underline{\mathds{1}}
      -
      \alpha_{\mathrm{B}}(\omega)
      \int\frac{\mathrm{d} q}{2\pi}~
      \underline{G}(q,\mathbf{R}_a,\omega_q^+)
      \right]^{-1}\alpha_{\mathrm{B}}(\omega)~.
  \end{align}
Here, $\alpha_{\mathrm{B}}(\omega)=\alpha_0\omega_a^2/  \left( \omega_a^2 - [\omega+\imath  0^{+}]^2 \right)$ 
is the (causal) bare polarizability
~\cite{Intravaia16}. 
We would like to note that the main difference between the bare and the 
dressed polarizability is the appearance of the Green tensor in 
Eq.~\eqref{Eq:PolDressed} which indicates that the particle dynamics 
is modified by its interaction with the electromagnetic environment. 
Given that $\underline{G}^{*}(q,\mathbf{R}_{a}, \omega)=\underline{G}(-q, \mathbf{R}_{a}, -\omega)$, 
the dressed polarizability fulfills the crossing relation $\underline{\alpha}^{*}(\omega,v)=\underline{\alpha}(-\omega,v)$.
Furthermore, upon utilizing the reciprocity of the involved materials, 
mathematically expressed by 
$\underline{G}^{\sf T}(q, \mathbf{R}_{a}, \omega)=\underline{G}(-q, \mathbf{R}_{a}, \omega)$,
we can derive with the help of Eq.~\eqref{Eq:PolDressed} that
$\underline{\alpha}(\omega,v)= \underline{\alpha}^{\sf T}(\omega,-v)$.
This result is in agreement with our general considerations. Indeed, in this case the linear polarizability does not depend on the
state of the system and coincides with the expression we would have obtained from the equivalent classical system.

We are now in the position to determine the atomic power spectrum  
$\underline{S}(\omega,v)$ from the relation
$
  \langle \hat{\mathbf{d}}(\omega)\hat{\mathbf{d}}^\mathsf{T}(\tilde{\omega})\rangle
  =4\pi^2 \delta(\omega+\tilde{\omega}) \underline{S}(\omega,v)
$
~\cite{Intravaia19a,Reiche20a}. 
Upon considering the stationary solution for the dipole's dynamics 
as given by Eq.~\eqref{dss} together with the properties of the 
polarizability and the expression for the field correlator in 
Eq.~\eqref{eq:efield}, we can write
  \label{Eqs:SpecPol}
  \begin{align}
    \label{Eq:FDTNEq}
    \underline{S}(\omega,v)
     & =
    \underline{\alpha}(\omega, v)
    \underline{\mathcal D}(\omega, v)
    \underline{\alpha}^{\dagger}(\omega, v).
  \end{align}
In this expression, we have introduced the tensor $  \underline{\mathcal D}(\omega, v)$ 
  \begin{align}
    \label{Eq:FieldSpectrum}
    \underline{\mathcal D}(\omega, v)
     & =
    \frac{\hbar}{\pi}
    \int \frac{\mathrm{d} q}{2\pi}
    \frac{ \underline{G}_{\Im}(q,\mathbf{R}_a,\omega_{q}^+)}{1-e^{-\beta\hbar\omega^{+}_{q}}}~.
  \end{align}

With the above results, we can explicitly check the general
properties of the power-spectrum tensor which we have discussed in Sec.~\ref{genprop} and at the beginning of Sec.~\ref{qfriction}.
To start, we notice that $ \underline{S}$ is Hermitian and in general not real-valued, 
as expected. It does not fulfill the FDT but, instead, fulfills Eq.~\eqref{noneqDef}. 
To see this more clearly, we consider the following identity which we can derive 
from Eq.~\eqref{Eq:PolDressed}
\begin{equation}
\label{alphaid}
\underline{\alpha}_{\Im}(\omega,v)=\int \frac{\mathrm{d} q}{2\pi}\;\underline{\alpha}(\omega,v) \underline{G}_{\Im}(q,\mathbf{R}_{a},\omega_{q}^+) \underline{\alpha}^{\dag}(\omega,v)~.
\end{equation}
In fact, this equation allows us to write an explicit expression for $\underline{J}(\omega)$ as
\begin{align}
\underline{J}(\omega)=
    \frac{\hbar}{2\pi}
    \int\frac{{\rm d} q}{2\pi}
& \left[\coth\left(\frac{\beta\hbar\omega^{+}_{q}}{2}\right)-\coth\left(\frac{\beta\hbar\omega}{2}\right)\right]
    \nonumber\\
    &\times\underline{\alpha}(\omega,v) \underline{G}_{\Im}(q,\mathbf{R}_{a},\omega_{q}^+) \underline{\alpha}^{\dag}(\omega,v)~.
\end{align}
Upon utilizing the properties of the polarizability and of the Green tensors mentioned above, we can show that, consistent with the discussion in Sec.~\ref{genprop}, $\underline{J}$ is Hermitian and fulfills the crossing relation. Moreover,
in earlier work with our co-workers, we have demonstrated within the above
framework that the local thermal equilibrium approximation displays significant
drawbacks regarding the quantitative evaluation of quantum friction. Specifically, neglecting $\underline{J}$, i.e. enforcing
the local thermal equilibrium approximation, underestimates the strength of the 
interaction and adversely affects the thermodynamical consistency of the 
physical system
~\cite{Intravaia16a,Intravaia19a,Reiche20c}.

Consider now the power-spectrum tensor. Noting that $(1-e^{x})^{-1}=1-(1-e^{-x})^{-1}$ and utilizing
the expressions in Eqs.~\eqref{propgi} and \eqref{alphaid} as well as the 
properties of the dressed polarizability, we find 
\begin{equation}
 \underline{S}(-\omega,v)= \underline{S}^{\sf T}(\omega,v)-\frac{\hbar}{\pi}\underline{\alpha}^{\sf T}_{\Im}(\omega,v)~.
\end{equation}
This coincides with Eq.~\eqref{reverse} which we have obtained in the first 
part of our analysis. 
We may conveniently check further properties once we specify the form for the Green tensor
$\underline{G}(q,\mathbf{R}_{a}, \omega)$. In general, the Green tensor can 
be decomposed into two terms. The first term, $\underline{G}^{(\rm v)}$, 
contains the geometry-independent vacuum behavior and has always the same form. 
Due to the homogeneity and isotropy of vacuum, $\underline{G}^{(\rm v)}(q,\mathbf{R}_{a}, \omega)$ 
does not depend on position, is symmetric and $\underline{G}_{\Im}^{(\rm v)}=\underline{G}_{I}^{(\rm v)}$ 
is diagonal~
\cite{Pieplow13,Amorim17}.
The second term, $\underline{G}^{(\rm sc)}$, describes the scattering and 
absorption due to the presence of material bodies and, consequently, depends 
on their composition, shape and arrangement.
As an example, we consider a half space, comprised of an isotropic material, located 
at $z<0$ and having a planar interface with vacuum at $z=0$. In this case, the 
mathematical expression for $\underline{G}^{(\rm sc)}$ can be found in the 
literature (see for example
~\cite{Tomas95,Pieplow13,Amorim17,Intravaia16}).
Due to the symmetry of the system, we may assume without loss of generality that the particle moves along the $x$-axis and choose
$\mathbf{R}_{a}\equiv(y_{a}=0,z_{a})$. Indeed, since the planar structure is 
translationally invariant along the $y$-axis, the choice of $y_{a}=0$ 
is physically irrelevant.  
If we denote the three-dimensional wave vector with $\mathbf{K}=(k_{x},k_{y},k_{z})$, 
we have $q\equiv k_{x}$ in Eq.~\eqref{totalForce}. In addition, we define 
$k=\sqrt{k_{x}^{2}+k_{y}^{2}}$ so that $k_{z}=-\imath\kappa=\sqrt{k^{2}-\omega^{2}/c^{2}}$ 
($\mathrm{Re}[\kappa]>0$ and $\mathrm{Im}[\kappa]<0$). Based on this,
we can write
\begin{equation}
\label{Gsc}
\underline{G}^{(\rm sc)}(k_{x},z_{a}, \omega)= \frac{\omega^{2}}{\epsilon_{0}c^{2}}\int \frac{d k_{y}}{2\pi}
\left(\underline{\Pi}_{p} r^{p}+\underline{\Pi}_{s}r^{s}\right)
\frac{e^{-2\kappa z_{a}}}{2\kappa} ,
\end{equation}
where $\epsilon_{0}$ is the vacuum permittivity and $r^{\sigma}\equiv r^{\sigma}(\omega,k)$ 
denote the polarization-dependent ($\sigma=s,p$) reflection coefficients 
of the surface~\cite{Wylie84}. The expressions for $r^{\sigma}$ depend on the specific properties of 
the material half-space but, due to its isotropy, they can only depend on $k$. 
Furthermore, in the above expressions, we have 
defined the polarization matrices $\underline{\Pi}_{p}$ and $\underline{\Pi}_{s}$, which, owing to the symmetries of the expression in Eq.~\eqref{Gsc}, can be written as
 \begin{align}
\underline{\Pi}_{s}=
\begin{pmatrix}
\frac{k_{y}^{2}}{k^{2}}&0&0\\
0&\frac{k_{x}^{2}}{k^{2}}&0\\
0&0&0
\end{pmatrix}, \quad
\underline{\Pi}_{p}= \frac{c^{2}\kappa^{2}}{\omega^{2}}
\begin{pmatrix}
\frac{k_{x}^{2}}{k^{2}}&0&-\imath \frac{k}{\kappa}\frac{k_{x}}{k}\\
0&\frac{k_{y}^{2}}{k^{2}}&0\\
\imath \frac{k}{\kappa}\frac{k_{x}}{k}&0&\frac{k^{2}}{\kappa^{2}}
\end{pmatrix} .
\end{align}

With the above, we can already check that the expression in Eq.~\eqref{Gsc} has 
a form that is consistent with Eq.~\eqref{split} where $c_{n}=0$ and 
$\mathbf{s}_{\perp}=s_{y}\mathbf{y}$. As discussed earlier, we may view
this as being the consequence of the mirror-symmetry of the total 
system with respect to the $y=0$ plane.
Finally, inserting in the previous equations the expression for the Green tensor given above and using Eq.~\eqref{Eq:FDTNEq}, one can show that $\underline{S}(\omega,v)$ is given by the sum of a diagonal and an anti-diagonal matrix. This is in agreement with our general considerations about the structure of the tensor based on the symmetry of the system. Moreover, we can verify that $\underline{S}(\omega,-v)\neq \underline{S}^{\sf T}(\omega,v)$, as we should expect in terms of our general result in Eq.~\eqref{transposeIdentity}. Notice that this result is obtained despite the fact that $\underline{\alpha}(\omega, -v)= \underline{\alpha}^{\sf T}(\omega,v)$ and that from Eq.~\eqref{Eq:FieldSpectrum} $\underline{\mathcal D}(\omega, -v)= \underline{\mathcal D}^{\sf T}(\omega,v)$.
We can further verify that
\begin{equation}
\underline{S}(\omega,-v)- \underline{S}^{\sf T}(\omega,v)=
\begin{pmatrix}
0&0& f(\omega,v)\\
0&0&0\\
f(\omega,v)&0& 0
\end{pmatrix}
\end{equation}
where, in agreement with the relation in Eq.~\eqref{transposeIdentity}, $f(\omega,v)$ is a real function odd in $v$. The previous expression also implicitly confirms that
the diagonal elements of the power-spectrum tensor are even in $v$, while the off-diagonal terms arising from $\boldsymbol{\Omega}_{\perp}=\Omega_{y}\mathbf{y}$ are odd in the same variable.

\section{Conclusions}
\label{conclusions}

In summary, we have discussed a number of relations involving the power-spectral 
density of a, in general, quantum system in an arbitrary steady state. This includes a large 
class of far-from-equilibrium scenarios as well as the equilibrium situation 
as a limiting case.
We have adopted a minimalistic viewpoint and,
besides stationarity, our analyses do not rely on any specific information 
about the microscopic dynamics of the system. Therefore, our results have broad validity, including systems with non-Markorvian 
as well as nonlinear dynamics. This is particularly relevant for quantum systems in which the Markov approximation may lead to inconsistent results~\cite{Talkner86,Intravaia16}. 

More precisely, we have focused on the steady state of a multidimensional 
(multivariate) stochastic process, where the dynamics is described in terms 
of a three-dimensional vector operator. 
Since the steady state may correspond to nonequilibrium dynamics for which 
the principle of detailed balance can be violated, close-to-equilibrium relations 
such as the fluctuation-dissipation theorem are not necessarily fulfilled. 
Nonetheless, we have been able to derive and classify a collection of properties that 
characterize the power spectrum and its connection to other physical quantities 
such as the system's susceptibility. When appropriate, we have linked them to the system's 
dynamics, its quantum behavior, the flow of energy within it and the role of dissipation, 
discussing their close-to-equilibrium counterparts.
This provides us with a solid basis to investigate and assess the validity of approaches such as the local thermal equilibrium 
approximation.

Once additional information about the system, such as symmetries and details
about its structure, is provided, we may obtain further properties and relations among the different quantities that characterize the system. 
As an illustration, we have analyzed the nonconservative viscous dynamics of a particle 
that moves in the quantum electromagnetic vacuum parallel to a 
translationally invariant arrangement of material bodies. When the drag 
felt by the particle during its motion
is balanced by an external driving force, the system reaches a nonequilibrium 
steady state that is characterized by a speed $v$.
Without specifying the internal microscopic dynamics of the moving object, 
we have obtained a collection of relations that characterize the power-spectrum 
tensor, its dependence on the velocity $v$ and its role in determining the 
frictional interaction at zero temperature (quantum friction). 
In addition, for this system, we have tested the general results described in 
the main part of the manuscript, in the context of a specific model which provides 
the microscopic internal dynamics of a moving atom. In particular, we have 
modeled the atom's dipole vector operator in terms of a three-dimensional harmonic oscillator and have
derived and examined the corresponding power-spectrum tensor as well as the related matrices. 

Our results can be useful in a large variety of systems and specifically for numerous 
fluctuation-induced phenomena, such as
the (quantum) Brownian motion~\cite{Grabert83,Hanggi05,Groblacher15}, radiative heat-transfer~\cite{Greffet02,Joulain05,Biehs21,Vazquez-Lozano24} the Casimir effect~\cite{Antezza05,Ingold09,Shen25} and the Casimir-Polder interaction~\cite{Scheel08,Buhmann08a,Laliotis21}, including 
their nonequilbrium variants such as quantum friction~\cite{Dedkov02a,Volokitin07,Milton16,Reiche22}.
Since often the information about the statistical properties of the system is or can be encoded 
in the power spectral density, knowing its general properties can be very helpful for understanding 
its behavior and to test thermodynamic consistency of its descriptions~\cite{Reiche20c}.

\appendix

\section{Positive definite matrices}
\label{positivity}
In this appendix, we elaborate on the reasons for why some of the matrices encountered in the main text are positive semidefinite or positive definite.
We work in the Heisenberg picture in which the operator are time-dependent and the states are not.
Let us consider first the power-spectrum tensor. For a generic complex vector $\mathbf{u}$ and 
any real-valued function $f(t)$ we can define the operator
\begin{equation}
\hat{b}_{\mathbf{u}}(t)=\int_{-\infty}^{\infty}dt_{1} f(t-t_{1})\mathbf{u}\cdot \mathbfh{d}(t_{1})~.
\end{equation}
Since $\langle \hat{b}^{\dag}_{\mathbf{u}}(t)\hat{b}_{\mathbf{u}}(t)\rangle$ is 
always nonnegative it follows that
\begin{equation}
\langle \hat{b}^{\dag}_{\mathbf{u}}(t)\hat{b}_{\mathbf{u}}(t)\rangle=\int_{-\infty}^{\infty} |f(\omega)|^{2}\mathbf{u}^{*}\cdot\underline{S}(\omega)\cdot\mathbf{u}\; d\omega\ge0~.
\end{equation}
Then, since $|f(\omega)|^{2}\ge 0$, we must have that 
\begin{equation}
\mathbf{u}^{*}\cdot\underline{S}(\omega)\cdot\mathbf{u}\ge 0 \quad \forall \mathbf{u}~\forall \omega~,
\end{equation}
which proves that $\underline{S}(\omega)$ is positive semidefinite.

For discussing the properties of $\underline{\alpha}_{\Im}(\omega)$ we consider the perturbed Hamiltonian $\hat{\tilde H}_{t}=\hat{H}_{t}-\mathbfh{d}\cdot \mathbf{e}(t)$, 
where $\mathbf{e}(t)$ denotes a small ($\abs{\mathbf{e}(t)}\ll 1$) real three-dimensional 
vector which is different from zero for a finite interval of time $\mathbf{e}(t\to \pm \infty)\to 0$. The perturbation introduces an additional intrinsic time-dependence
with respect to that possibly characterizing the unperturbed Hamiltonian. 
In linear order of perturbation theory, the expectation value of the associated time 
evolution in the Heisenberg picture of the dipole operator as prescribed by $\hat{\tilde H}_{t}$ (see also below), i.e. $\hat{\tilde{\mathbf{d}}}(t)$, can be written as
\begin{align}
\label{linearresp}
\langle\hat{\tilde{\mathbf{d}}}(t)\rangle\equiv \mathrm{tr}[\hat{\tilde{\mathbf{d}}}(t)\hat{\rho}]
&\stackrel{\abs{\mathbf{e}(t)}\ll 1}{\approx}
\int_{-\infty}^{\infty}\mathrm{d}t_{1}\;\underline{\alpha}(t-t_{1}) \cdot \mathbf{e}(t_{1})~,
\end{align}
where we used the assumption that $\langle\hat{\mathbf{d}}(t)\rangle=0$ (see Sec. \ref{genprop}) and $\underline{\alpha}(\tau)$ is the linear response tensor with elements defined in Eq.~\eqref{polar}.
Mathematically, we may regard the tensor as the functional derivative with respect 
to the external perturbation of the exact time evolution, i.e.
$\delta\langle\hat{\tilde d}_{i}(t)\rangle/\delta e_{j}(t_{1})= \alpha_{ij}(t-t_{1})$.

In general, the time evolution according to $\hat{\tilde H}_{t}$ of an operator $\hat{A}_{t}$ which intrinsically depends on 
	      time is given by $\hat{\tilde A}(t)=\hat{\tilde{U}}^{\dag}_{t}\hat{A}_{t}\hat{\tilde{U}}_{t}$,. The operator $\hat{\tilde{U}}_{t}$ is the solution of  $\partial_{t}\hat{\tilde{U}}_{t}=(\imath/\hbar)\hat{\tilde H}_{t}\hat{\tilde{U}}_{t}$
	      with the initial condition $\hat{\tilde{U}}_{t=0}= \underline{\mathds{1}}$.
Accordingly, the total time derivative of $\hat{\tilde A}_{t}(t)$ is given by	      
\begin{align}
\frac{d}{dt}\hat{\tilde A}(t)=\frac{\imath}{\hbar}\hat{\tilde{U}}^{\dag}_{t}[\hat{\tilde H}_{t},\hat{A}_{t}]\hat{\tilde{U}}_{t}+\hat{\tilde{U}}^{\dag}_{t}(\partial_{t}\hat{A}_{t})\hat{\tilde{U}}_{t}~.
\end{align}	      
We notice that for $\hat{A}_{t}\equiv \hat{\tilde H}_{t}$ the first term on the r.h.s. of the previous equation is identically zero.
As a consequence, during the perturbation, the total change of energy is given by
\begin{align}
\int_{-\infty}^{\infty} dt\; \frac{d}{dt}\langle \hat{\tilde H}_{t}(t)\rangle
&=\int_{-\infty}^{\infty} dt\; \langle \frac{d}{dt} \hat{\tilde H}_{t}(t)\rangle
=\int_{-\infty}^{\infty} dt\; \langle U^{\dag}_{t}(\partial_{t}\hat{\tilde H}_{t})U_{t}\rangle
\nonumber\\
&=\int_{-\infty}^{\infty} dt\; \langle U^{\dag}_{t}(\partial_{t}\hat{H}_{t})U_{t}\rangle
-\int_{-\infty}^{\infty}\mathrm{d}t\; \langle\hat{\tilde{\mathbf{d}}}(t)\rangle \cdot \frac{d\mathbf{e}(t)}{dt}~.
\end{align}
While the first term in the last line of the previous expression gives the change in energy due to the intrinsic time dependence of the unperturbed Hamiltonian $\hat{H}_{t}$, the second term describes the modification of the energy due to the perturbation alone.
Inserting Eq.~\eqref{linearresp} we have that
\begin{equation}
-\int_{-\infty}^{\infty}\mathrm{d}t\; \langle\hat{\tilde{\mathbf{d}}}(t)\rangle \cdot \frac{d\mathbf{e}(t)}{dt}
=\int_{0}^{\infty}\frac{\mathrm{d}\omega}{\pi}\;\omega\; \mathbf{e}^{*}(\omega)\cdot\underline{\alpha}_{\Im}(\omega)\cdot\mathbf{e}(\omega)~,
\end{equation}
where $\underline{\alpha}_{\Im}(\omega)$ is the the matrix defined in Eq.~\eqref{alphastrangei} of the main text. 
If we require that the change of energy induced by 
the perturbation is always positive irrespective of the form of the perturbation, 
then for $\omega>0$ the matrix $\underline{\alpha}_{\Im}(\omega)$ must be 
positive definite. Physically, this is equivalent to the request that, despite its intrinsic and possibly externally driven dynamics, the system remains stable and is able to dissipate the energy arising from the perturbation.
Moreover, since~\cite{Landau80a}
\begin{equation}
\underline{\alpha}(\imath\xi)=\frac{1}{\imath\pi}\int_{-\infty}^{\infty}d\omega \frac{\omega}{\omega^{2}+\xi^{2}}\underline{\alpha}(\omega)
\end{equation}
we can write that
\begin{equation}
\frac{\underline{\alpha}(\imath\xi)+\underline{\alpha}^{\sf T}(\imath\xi)}{2}=\frac{1}{\pi}\int_{-\infty}^{\infty}d\omega \frac{\omega}{\omega^{2}+\xi^{2}}
\underline{\alpha}_{\Im}(\omega)~.
\end{equation}
Given that the real part of a positive definite matrix is also positive definite and that the real and imaginary part of $\underline{\alpha}_{\Im}(\omega)$ are, respectively, odd and even in $\omega$, we obtain the expression in Eq.~\eqref{alphasymm} of the main text. It indicates that, if $\underline{\alpha}_{\Im}(\omega)$ is 
positive definite, the symmetric part of $\underline{\alpha}(\imath\xi)$ for $\xi\ge 0$ is also positive definite.

Finally, the previous considerations allow us to say that the matrix $\underline{\nu}(\omega)$ defined in Eq.~\eqref{defnu} is also positive semidefinite \cite{Fleming13}. Indeed, using its definition and properties discussed in the main text, we have that
\begin{equation}
\underline{\nu}(\omega)=\underline{\nu}^{*}(-\omega)=\underline{S}^{*}(-\omega)+\frac{\hbar}{2\pi}\underline{\alpha}_{\Im}(\omega)~,
\end{equation} 
or also combining this again with Eq.~\eqref{defnu} that
\begin{equation}
\underline{\nu}(\omega)=\frac{\underline{S}(\omega)+\underline{S}^{*}(-\omega)}{2}=\frac{\underline{S}(\omega)+\underline{S}^{\sf T}(-\omega)}{2}~.
\end{equation} 
Both expression show that $\underline{\nu}(\omega)$ for $\omega\ge 0$ is positive semidefinite because it is given by the sum of positive semidefinite matrices.

\section{Symmetry properties}
\label{appSymm}
In this appendix, we provide additional information about the 
consequences of the mirror-symmetry on the frictional quantum system discussed 
in Sec.~\ref{qfriction}. \\
We start by considering the tensor $\underline{G}_{\Im}(q,\mathbf{R}_{a}, \omega)$. 
Since it is Hermitian, we can write 
\begin{align}
\label{dec1}
\underline{G}_{\Im}(q,\mathbf{R}_{a}, \omega)=&\underline{D}(q,\mathbf{R}_{a}, \omega)
\nonumber\\
&+\mathbf{c}(q,\mathbf{R}_{a}, \omega)\cdot\mathbf{\underline{\Lambda}}+\mathbf{s}(q,\mathbf{R}_{a}, \omega)\cdot\mathbf{\underline{L}}
\nonumber\\
=&\underline{D}(q,\mathbf{R}_{a}, \omega)
\nonumber\\
&+c_{n}(q,\mathbf{R}_{a}, \omega)\underline{\Lambda}_{n}+\mathbf{c}_{\perp}(q,\mathbf{R}_{a}, \omega)\cdot\mathbf{\underline{\Lambda}}_{\perp}
\nonumber\\
&+s_{n}(q,\mathbf{R}_{a}, \omega)\underline{L}_{n}+\mathbf{s}_{\perp}(q,\mathbf{R}_{a}, \omega)\cdot\mathbf{\underline{L}}_{\perp}~,
\end{align}
where $\underline{D}(q,\mathbf{R}_{a}, \omega)$ is a diagonal matrix, 
$\mathbf{c}(q,\mathbf{R}_{a}, \omega)$ and $\mathbf{s}(q,\mathbf{R}_{a}, \omega)$ 
are two real-valued three-dimensional vectors. The matrices $\underline{\Lambda}_{i}$ 
and $\underline{L}_{i}$ are  defined in the main text (see the text 
below Eqs.~\eqref{decompositionS} and \eqref{decompositionK}). 
The subscript $n$ and $\perp$, respectively, indicate the components 
parallel and orthogonal to the direction of translational invariance 
that is given by $\mathbf{n}$.
Since this decomposition is unique, we can utilize Eq.~\eqref{trps} and the properties of
the involved matrices
to conclude that $\underline{D}$ and $\mathbf{c}$ are even functions  
of $q$, while $\mathbf{s}$ is an odd function of $q$.

If we consider the symmetry described by $\underline{M}_{n}=\underline{\mathds{1}}-2\mathbf{n}\mathbf{n}^{\sf T}$, Eq.~\eqref{symmG} 
reveals that for the setup under consideration the above expression must 
coincide with
\begin{align}
\label{dec2}
\underline{M}_{n}\underline{G}_{\Im}&(-q,\mathbf{R}_{a}, \omega)\underline{M}^{-1}_{n}=\underline{D}(-q,\mathbf{R}_{a}, \omega)
\nonumber\\
&+c_{n}(-q,\mathbf{R}_{a}, \omega)\underline{\Lambda}_{n}-\mathbf{c}_{\perp}(-q,\mathbf{R}_{a}, \omega)\cdot\mathbf{\underline{\Lambda}}_{\perp}
\nonumber\\
&+s_{n}(-q,\mathbf{R}_{a}, \omega)\underline{L}_{n}-\mathbf{s}_{\perp}(-q,\mathbf{R}_{a}, \omega)\cdot\mathbf{\underline{L}}_{\perp}~.
\end{align}
Here, we have used that diagonal matrices and $\underline{\Lambda}_n$ 
are invariant with respect to the transformation while for the other 
matrices we have 
$\underline{M}_{n}\boldsymbol{\underline{\Lambda}}_{\perp}\underline{M}^{-1}_{n}=-\boldsymbol{\underline{\Lambda}}_{\perp}$ 
and 
$\underline{M}_{n}\mathbf{\underline{L}}_{\perp}\underline{M}^{-1}_{n}=-\mathbf{\underline{L}}_{\perp}$.
When comparing the coefficients of Eqs.~\eqref{dec1} and \eqref{dec2}, 
we obtain 
$\mathbf{c}_{\perp}(-q,\mathbf{R}_{a}, \omega) = - \mathbf{c}_{\perp}(q,\mathbf{R}_{a}, \omega)$ 
(odd function in $q$) and 
$s_{n}(q,\mathbf{R}_{a}, \omega) = s_{n}(-q,\mathbf{R}_{a}, \omega)$ 
(even function in $q$). 
This is, however, in contradiction with the the constraints given by 
Eq.~\eqref{trps} and, consequently, these expressions must vanish.  
This leads to the decomposition in Eq.~\eqref{split}. 
We can follow a similar procedure if the system features an additional 
symmetry $\underline{M}_{t}$, where $\mathbf{t}\cdot\mathbf{n}=0$. 
If the symmetry plane contains both, $\mathbf{q}=q\mathbf{n}$ and 
$\mathbf{R}_{a}$, thus making them invariant with respect to $\underline{M}_{t}$,
we must have that $\underline{G}_{\Im}$ in Eq.~\eqref{split} is invariant 
with respect to this transformation. 
However, if we call $\mathbf{t'}=\mathbf{n}\times \mathbf{t}$, we obtain
that
\begin{align}
\underline{G}_{\Im}&=\underline{D}+c_{n}\underline{\Lambda}_{n}+s_{t'}\underline{L}_{t'}+s_{t}\underline{L}_{t}
\nonumber\\
&=\underline{M}_{t}\underline{G}_{\Im}\underline{M}_{t}^{-1}=\underline{D}-c_{n}\underline{\Lambda}_{n}-s_{t'}\underline{L}_{t'}+s_{t}\underline{L}_{t},
\end{align}
which, in turn, implies that $c_{n}=s_{t'}=0$.

The power-spectrum tensor $\underline{S}(\omega,v)$ is 
subject to fewer restrictions. In general, we have that
\begin{align}
\underline{S}(\omega,v)&=\underline{\Delta}(\omega,v)
+\boldsymbol{\chi}(\omega,v)\cdot\mathbf{\underline{\Lambda}}+\boldsymbol{\Omega}(\omega,v)\cdot\mathbf{\underline{L}}
\nonumber\\
&=\underline{\Delta}(\omega,v)
+\chi_{n}(\omega,v)\underline{\Lambda}_{n}+\boldsymbol{\chi}_{\perp}(\omega,v)\cdot\mathbf{\underline{\Lambda}}_{\perp}
+\Omega_{n}(\omega,v)\underline{L}_{n}+\boldsymbol{\Omega}_{\perp}(\omega,v)\cdot\mathbf{\underline{L}}_{\perp}~.
\end{align}	
Due to the symmetry of the system and its dissipative nature, we can 
anticipate that $\Omega_{n}=0$. Arguments analogous to those made for 
$\underline{G}_{\Im}$ reveal that $\underline{\Delta}$ and $\chi_{n}$ 
are even functions of the velocity $v$, while $\boldsymbol{\chi}_{\perp}$ 
and $\boldsymbol{\Omega}_{\perp}$ must be a odd function of $v$. Therefore,
we can write
\begin{align}
\underline{S}^{\sf T}(\omega,-v)&=\underline{\Delta}^{\sf T}(\omega,-v)
+\boldsymbol{\chi}(\omega,-v)\cdot\mathbf{\underline{\Lambda}}^{\sf T}
+\boldsymbol{\Omega}_{\perp}(\omega,-v)\cdot\mathbf{\underline{L}}_{\perp}^{\sf T}
\nonumber\\
&=\underline{\Delta}(\omega,v)+\chi_{n}(\omega,v)\underline{\Lambda}_{n}-\boldsymbol{\chi}_{\perp}(\omega,v)\cdot\mathbf{\underline{\Lambda}}_{\perp}
+\boldsymbol{\Omega}_{\perp}(\omega,v)\cdot\mathbf{\underline{L}}_{\perp}
\nonumber\\
&=\underline{S}(\omega,v)-2\boldsymbol{\chi}_{\perp}(\omega,v)\cdot\mathbf{\underline{\Lambda}}_{\perp}~,
\end{align} 
which is equivalent to Eq.~\eqref{transposeIdentity}. Again, when an 
additional mirror-symmetry $\underline{M}_{t}$ occurs and $\mathbf{v}=v\mathbf{n}$ 
and $\mathbf{R}_{a}$ are contained in (or parallel to) the symmetry plane, then 
$\underline{S}(\omega,v)$ must be invariant with respect to the transformation. 
With arguments analogous to the above-discussed case of $\underline{G}_{\Im}$, 
we can then conclude that $\boldsymbol{\chi}=\chi_{t}\mathbf{t}$ and 
$\boldsymbol{\Omega}_{\perp}=\Omega_{t}\mathbf{t}$.

\ack{}

We sincerely thank Bettina Beverungen, Daniel Reiche and Marty 
Oelschl{\"a}ger for numerous motivating questions and valuable 
discussions around many of the results presented here.

%\section*{References}

%\bibliographystyle{iopart-num}
%\bibliography{/Users/nabu/Documents/Lavoro/bibliography/biblio.bib}

\providecommand{\newblock}{}

\end{document}